# Some comments on events associated with falling terrestrial rocks and iron from the sky

Andrei Ol'khovatov

https://orcid.org/0000-0002-6043-9205

Independent researcher

Russia, Moscow

email: olkhov@mail.ru

**Abstract.** Some examples of the events associated with falls of rocks and iron terrestrial origin from the sky are considered (in scientific publications they are often called as meteor-wrongs or pseudo-meteorites). Their possible connections with other natural phenomena (like a whirlwind and a ball-lightning) are considered. Some compilation of info connected with the event in the US town of Elma in 2003 is presented. Also some possible parallels with the 1908 Tunguska event are given. The author hopes to draw attention to the need for more detailed research of such phenomena.

## 1. Introduction

There are many reports of a rock falling from the sky, but later discovered to be of terrestrial origin (nowadays they are often called as meteor-wrongs or pseudo-meteorites). Info on such events was published in scientific journals in 18-19 centuries, but rarely in nowadays. It is not because the events stopped to occur, but because modern science is rarely interested in them. However there are fortunately some exceptions, like US scientist Randy L. Korotev ( https://sites.wustl.edu/meteoritesite/people/randy-l-korotev/ ), and Spanish scientist Jesús Martínez Frías ( http://tierra.rediris.es/jmfrias/ ). From the Korotev's web-page it is seen that such objects continue to fall nowadays: https://sites.wustl.edu/meteoritesite/items/thud/

The author of this article is especially interested in events which were accompanied by fireball--like phenomena, and collects info about such events for many years placing the info on web-page http://olkhov.narod.ru/gr1997.htm .

Let's start with the July 15, 2003 event in the US town of Elma (Wa.). The author looked after the event soon after it occurred and contacted some people who were involved. Important info was got from Tami Hickle (Elma High School), and internet forum Meteorite Mailing List (http://www.meteorite-list-archives.com/ ). Also local mass-media (for example, a newspaper "The Daily World", KOMO TV) delivered some info. Unfortunately some the data already vanished from internet. To the author knowledge, no info on this remarkable event was published in scientific literature. So the main goal of this article is to preserve the info on the event for future science. Maybe the article will urge some researchers to continue research of this remarkable event. This research was private research of the author.

## 2. The Elma event

Here is a summary of the 2003 Elma event combined from various sources. The reliability of the sources is different, so in some unclear cases the word "seems" is added.

Three primary eyewitness (young men) were driving at night, and saw a big, bright flash. At first, it was thought that it was a shooting star. According to the witnesses it had a tail about six or seven - feet long (as they thought). Then they watched as the fireball disappeared in the direction of their high school and after this saw a cloud of dust go up in the air.

The eyewitness decided to investigate, and drove toward Elma High School, where they thought the fragments fell. The school's security cameras did not detected flash of light. The above-mentioned witnesses were seen on the cameras driving up and going into the field to look around at about 12:34 a.m. July 15. Then they left and returned with the police about 12:59 a.m. (T. Hickle, personal communication, July, 2003). From this info it looks like the most accurate time estimation of the event is about 7.30 UTC (July 15, 2003).

The witnesses said that golf ball size holes in the sand were with the glassy black rocks they picked up near the school at the school athletic fields. They said the rocks were hot, and one of them even burned a thumb and finger. Police joined the young men, examining asphalt and the ground.

KOMO 4 News also received many e-mails from other independent viewers saying they too saw the fireball streak across the sky shortly before 1 a.m.

Soon other people arrived to search after hearing the story. Amateur meteorite hunters as well as scientists were presented too. Analysis of the discovered stones showed that they are not meteorites.

Tami Hickle from the Elma High School informed me in July 2003 that the object passed over Elma headed north. One of the reasons for this is a man who lives about a mile north of the school site - reported seeing the flash going over his place and reports that it continued north.

T. Hickle also emailed me in July 2003 that there was a news article at web-site www.thevidette.com for the date of July 17, 2003. A lady from Montesano (10 miles west of Elma) says she was awakened by what sounded like falling gravel the same night. T. Hickle sent me the following details. The Vidette had this report: Helen Schweitzer of Montesano, (10 miles
west of Elma) said that at 4:33 this morning (July 15) she woke up. It was like somebody was dropping a half load of gravel in her driveway. Except it sounded like it was coming from underneath her house. Her dog was panicky. Later that morning, Schweitzer noticed a small rock on her driveway.

The author did not find any info on the development of the Montesano story. Anyway the fallen object is absent in the Meteoritical Bulletin Database which has been constructed and is maintained by the Nomenclature Committee of the Meteoritical Society ( https://www.lpi.usra.edu/meteor/about.php ).

Unfortunately no info on the "Elma stones" composition was published in scientific publications. Some "non-official" info was presented in Meteorite Mailing List internet forum ( nowadays internet address http://meteorite-identification.com/mwnews/elma_tests.html ) by a prominent meteorite collector and hunter Adam Hupe (Greg Hupe took part in some of the posts too) who was in contact with researchers. Here some extracts from his forum's posts in 2003.

"- *...this material consist of sand incased in a basaltic glass shell. The materials in the*

*glass are not separated into swirls so what ever created these objects had a huge amount of energy and the glass was quenched very rapidly.*

*- The problem with this material is that it does not contain any metal. If it was in a reducing environment why isn't there any metal, even at microprobe levels? This almost certainly eliminates this material coming from an industrial process, at least none that I have ever heard of. If it came from a foundry why wasn't the sand melted in the center. The sand is also trapped in sealed vesicles, that look almost like chondrules, like a condensation product, which is extremely odd. It was extremely difficult to make thin-sections of this stuff because of the friability of the material trapped inside.*

*- Some of the things that I did was to collect samples from the supplier of the original sinter track material and to track down the source of the sand found in the shot-put pit.*

*The sand in the shot-put pit came from a quarry about a quarter of a mile away. It was determined that the sand is natural, clean and that the glassy objects were introduced sometime later.*

*Since sinters could be a possible explanation we decided it would be proper to compare them with what was found in the pit. The sinters are left over material from burning coal in a steam plant. The sinters in the original track have been covered for over a decade by a layer of asphalt and a covering of rubber. Since it was thought that these glassy objects might have been contamination from the original track it was important to locate the source of these sinters for comparison. Here is the problem, coal*

*varies from one burn to the next which means there will be some variance in the sinters. All I can say is that the sinters that were tested at the UW did not match the glassy objects found in the pit. The UW will release a public opinion, I believe, on the results tomorrow.*

*Things that we have no explanation for:*

*A dust cloud was observed from the pit within seconds of seeing the fireball.*

*An eyewitnesses finger and thumb were burned after picking up a piece.*

*A piece of this glass was found by a police officer embedded in a telephone pole next to the pit.*

*The same police officer reported that there were dents and burn marks on some nearby bleachers.*

*The same officer reported that where these black objects were found on an asphalt sidewalk that there was melting.*

*A newspaper photographer took a picture of hundreds of little craters in the sand, each containing these black glass objects at the bottom.*

*Sand which was not melted was found in the centers of most of these objects which consist of basaltic glass that was quenched very rapidly, kind of like a fulgurite.*

*The track coach claims the sand was clean prior to this event and sinters would never have been added.*

*A few institutions have weighed in on this event. One DOD contractor thinks it could have been a transient event (a plasma discharge) which they sporadically pick up on their sensors when monitoring for nuclear activity from time to time. A Russian scientist is convinced that it might have been a Geometeorite event (ball lightning). I think that the UW will report that yet more study needs to be done in order to connect this strange glass to the meteor sighting which was witnessed by several dozen people.*

*I will post microprobe pictures and data when the UW gives me the green light."*

Unfortunately the author fails to find any public opinion from researchers in the University of Washington on the Elma event.

In 2004 one more post appeared on the Elma event by Adam and Greg Hupe ( ( http://www.meteorite-identification.com/mwnews/elma.html ):

*"The Elma event was very interesting from an investigative standpoint. For us, it was more like investigating an X-File scene than a meteorite fall. Dozens of witnesses came forward including officer Bealert who also saw the light show and later investigated the suspected impact site. There is no doubt that an event resembling a meteor was observed that night. The thing in question is whether the material found behind the school in the shot-put pit is meteoritic or not. We have stated from the beginning that the recovered glassy objects were not meteorites.*

*Another theory arose soon after the material was determined not to be meteoritic. Russian researcher Dr. A. Ol'khovatov stated the conditions were perfect for a transient event and witness statements could be explained by such an occurrence. He went as far as documenting the event on his web-site calling it a probable a "Geometeorite". His interpretation can be found by sifting through a huge amount of data on the web-site below:*

*http://olkhov.narod.ru/gr1997.htm*

*Sandia National Laboratories has a division that investigates transient events and also believes the witness statements are consistent. They have inquired a few times and take these kinds of reports seriously. From what we were lead to believe they have the ability to track such events from remote sensing equipment.*

*The University of Washington analyzed the material and found in simple terms glass surrounding unmelted sand, kind of like a reverse fulgurate. We found the company who supplied the sand and investigated. The glassy objects had to have been introduced into the sand at a latter time than when supplied. It was suggested that sinters from the original nearby running track may have contaminated the shot-put pit. This seemed impossible because the running track had been rubberized over a decade ago and there were no glass objects found between the track and the pit. The source of the sinters was tracked down to a power generation plant. Samples were brought back and analyzed. The problem was the sinters were highly variable but none were found with unmelted sandy cores. They were somewhat similar leading the University of Washington to conclude they were probably the most likely answer to the puzzle.*

*An interesting item is that Robert Haag investigated a similar event as noted at the*

*site below:*

*http://www.100megsfree4.com/farshores/nrock03.htm*

*In conclusion two labs are leaning towards a transient event and one against. We will continue to keep an open mind but do not have much spare time to investigate anything other than true meteorites."*

The author did not find any later factual info on the Elma event at the forum.

Charles R. Viau from USA sent me several photos of the Elma fragments which are reproduced (with his kind permission) below as Fig.1-3.

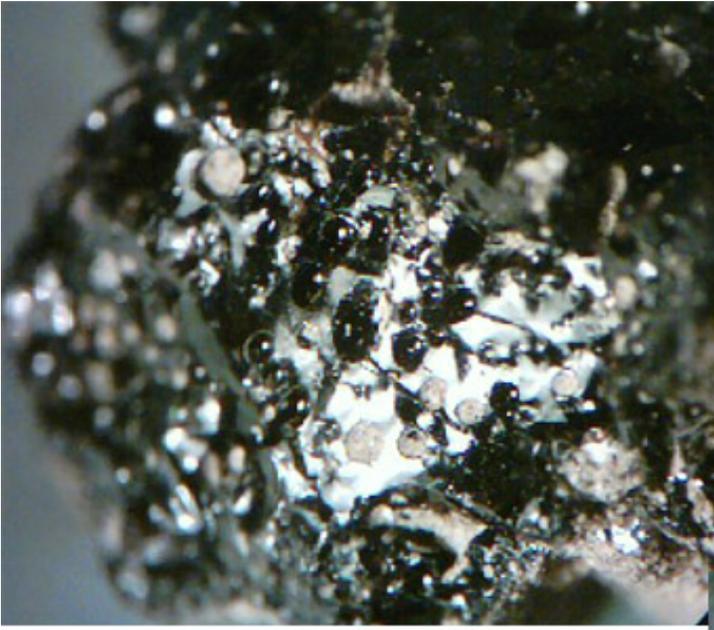
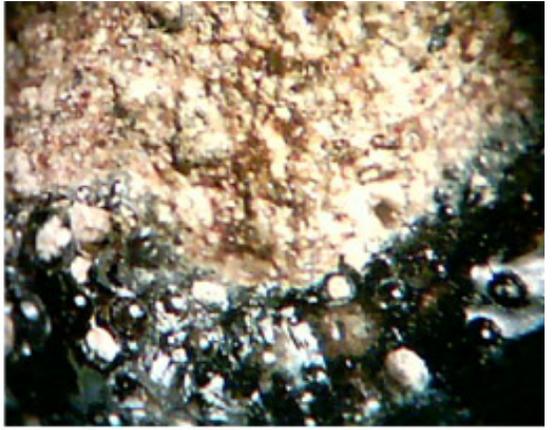
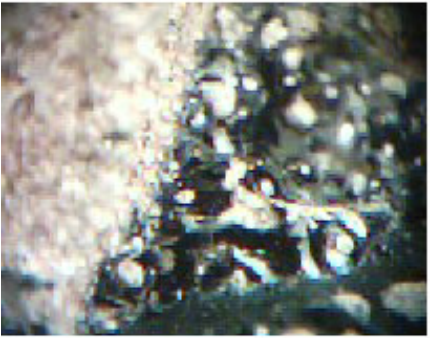
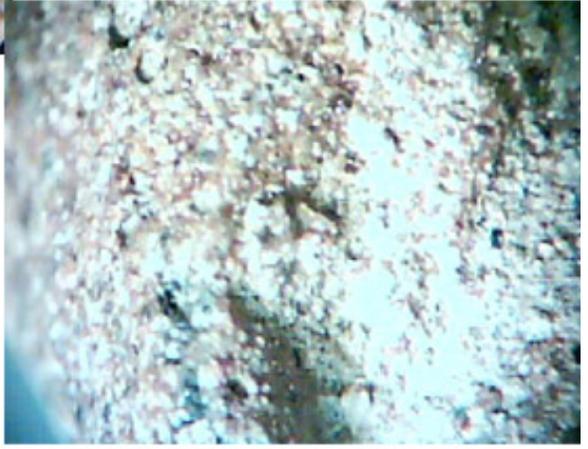
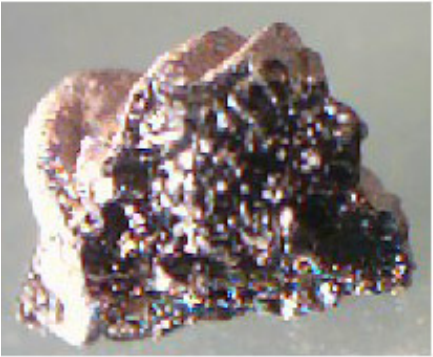
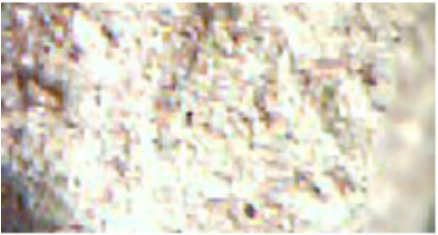
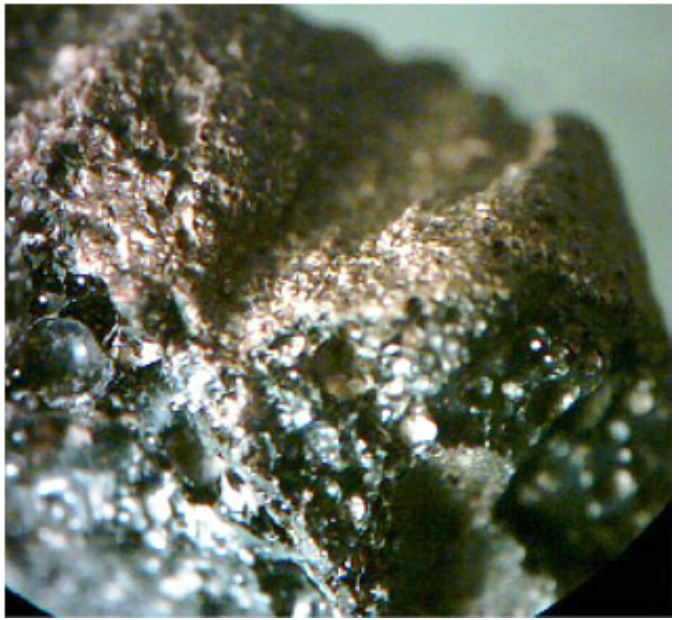

**Fig.1**

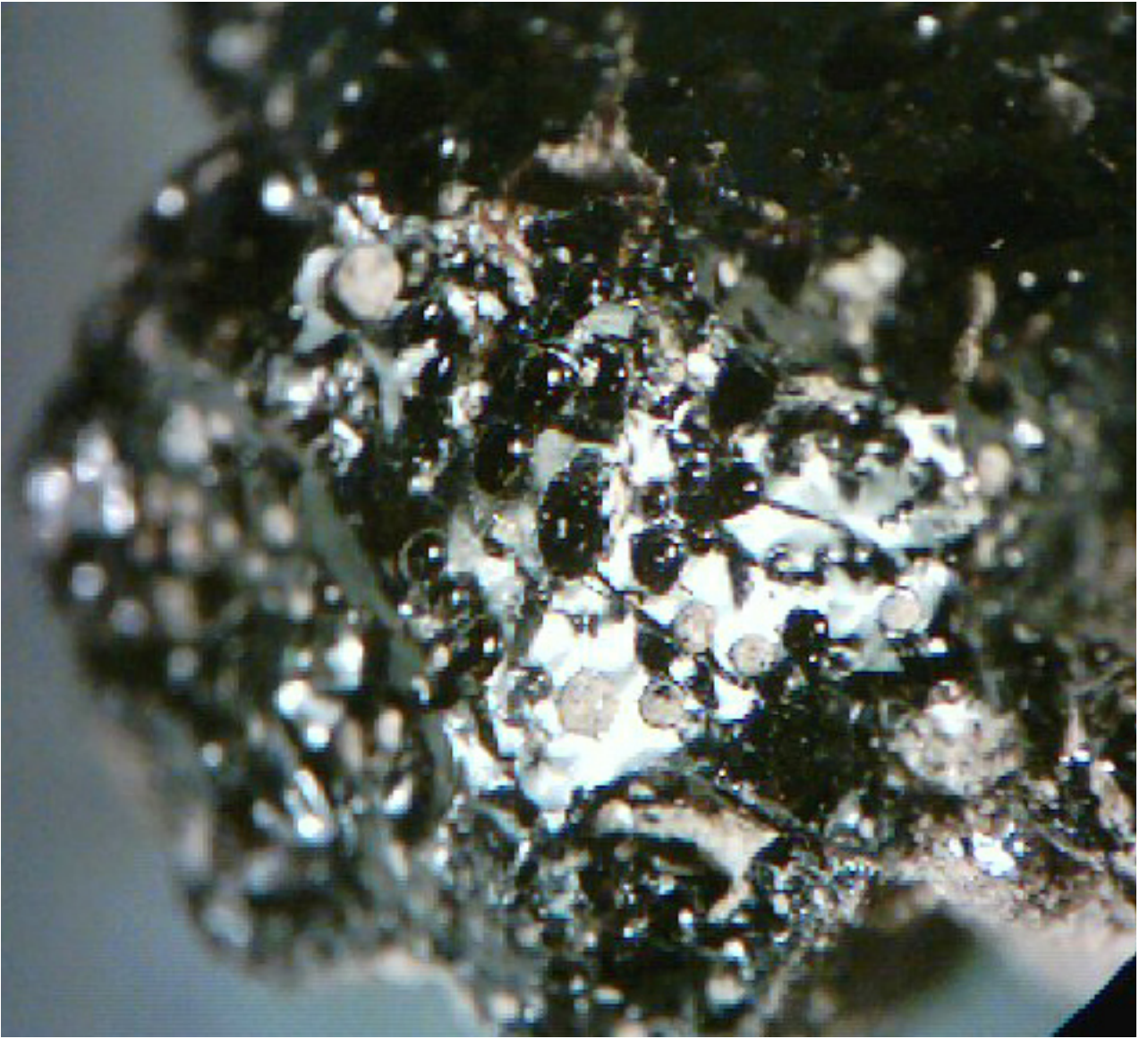

**Fig.2**

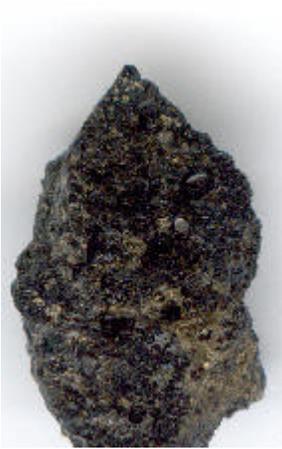

**Fig.3**

## 3. Discussion

The data points that the Elma event occurred at about 7.30 UTC, July 15, 2003. Several aspects attract attention in the event.

- The event was associated with the fireball.

- The discovered fragments were very hot.

- The area of fall was very compact (compared with typical spacebody infall events).

Also it looks like the area (where the fireball was seen) was rather small too (compared with typical spacebody infall events). These points that the altitude of the fireball was very small compared with spacebody infall events. These aspects exclude a possibility that the event was caused by entering the atmosphere by some known type of a spacebody (natural or man-made).

Unfortunately a physical mechanism of the event is unknown, so let's look whether similar events occurred. And indeed, the most often similar events were reported from time-to time in association with thunderstorms.

The stories about falling matter in association with thunderstorms are known from ancient times. Despite that some of them maybe are looking as dubious, however F. Arago ( who was very interested in thunderstorms and lightning) wrote (Arago, 1855):

*"Without certainly wishing to resuscitate superannuated ideas respecting "stone thunderbolts," I would remark that it is not proved that we ought to reject as false all the accounts which speak of strokes of lightning accompanied by the fall of material substances."*

Arago gave also the following example of the falling matter:

*"In July, 1681, near Cape Cod, the English ship 'Albemarle' was much damaged by lightning. The stroke of lightning was followed by a bituminous substance, which burnt with a smell resembling that of gunpowder, falling into the stern boat. It burnt itself out where it fell, after vain attempts had been made either to extinguish it with water, or to throw it overboard with wooden poles."*

A famous French astronomer and writer Camille Flammarion was also interested in such phenomena. Here is just one example from (Flammarion, 1905) which occurred during a powerful thunderstorm with hail:

*"On April 24, 1887, a storm burst over Mortrée (Orne), and the lightning literally chopped the telegraph wire on the route to Argentan for a distance of 150 yards. The pieces were so calcinated that they might have been under the fire of a forge; some of*

*the longer ones were bent and their sections welded together. The lightning entered by the door of a stable in the form of a fireball, and came near a person who was preparing to milk a cow;... <...>*

*The inexplicable phenomenon was that at the precise moment when the lightning crossed the stable,*
*a great quantity of incandescent stones fell before a neighbouring house. "Some of these fragments,*
*of the size of nuts," wrote the Minister of Post and Telegraphs at the Academy, "are of a not very*
*thick material, of a greyish-white, and easily broken by the fingers, giving forth a characteristic odour of sulphur. The others, which are smaller, are exactly like coke."*

At the beginning of the 20th century reports of the stones fall associated with thunderstorms still continued to be published in scientific literature, like, for example, the following one in the Nature journal (Bullen, 1912). During a heavy thunderstorm the observer stated that the stone fell within a few feet from where he was standing, and that it entered the ground for a distance of about 3 ft. Its fall was accompanied by an unusually heavy clap of thunder. The stone weighed about 2.5 kg, and was irregularly ovate on the one side, and broken in outline on the other. "The actual surface throughout was fairly deeply pitted, and under magnification exhibits the usual chondritic structure of the crystalline matter with interspersed particles of what appears to be nickeliferous iron".

However in the following March 21 issue of Nature G. E. Bullen wrote that the following detailed examination of the stone reveals that it is not a meteorite. Also the fallen object is absent in the Meteoritical Bulletin Database which has been constructed and is maintained by the Nomenclature Committee of the Meteoritical

Society ( https://www.lpi.usra.edu/meteor/about.php ).

Regarding a possible way of the stone transportation, it can be said that, for example, a ball-lightning can transport some substance. At least there is a known case when ball lightning left behind small metal balls. Let us describe this case based on the data (Gromyko, 2004a, 2004b).

The event took place on July 19, 2003. A ball of fire, the size of a soccer ball, floated into a room on the second floor. A few seconds later, an explosion was heard and red-hot metal balls with an average diameter of about 1 mm fell on the floor. Glowing balls quickly darkened, leaving burn marks on the floor linoleum. Most of the cooled balls were collected by eyewitnesses and analyzed at the Institute of Physics (Siberian Branch of Russian Academy of Science) in Krasnoyarsk. It was found that the balls are hollow spheres of pure iron, 0.3-1.2 mm in diameter, with a wall thickness of about 10 microns. The results of comparing the magnetization of these balls and α·Fe show that the shape of the curves is identical. There is no hysteresis in the materials under study. The graph of the spectrogram of stimulated Raman scattering (no resonance lines) confirms the amorphousness of the
ball-lightning samples.

There is also an article (Chen et al., 2014) where spectral analysis indicates that the radiation from soil elements is present for the entire lifetime of a detected ball-lightning.

It can be speculated that a kind of a whirlwind can responsible for transportation of objects in such events. There is at least one hint for such possibility. Here is an interesting observation reported in Nature [Worth, 1972]. A witness was standing on top of the Puy Mary, 1787 metres high, the highest of the volcanic cones in the range of the Cantal mountains. To the north, about 1-2 miles away, he saw lightning coming out from a black bank of clouds and heard the thunder about 5-6 s later. However, he felt a fairly strong blast of hot air reaching him before he could hear the thunderclap. This happened several times in quick succession. The blast was almost silent.

Interestingly that some "hot" effects can occur without a thunderstorm (but maybe associated with some cloud, especially with a dark one). Here is an example (Anonymous, 1857). A witness reported about a fall near Ottawa, Illinois specimens of "scoria" on June 17, 1857. The time it occurred was fifteen or twenty minutes before 2 p. m.; the wind blew west by south. The witness said that the cinders fell in a northeasterly direction in the shape of the letter V. The weather had been showery, but the witness heard no thunder and saw no lightning. There appeared to be a small, dense black cloud hanging over the garden in a westerly direction, or a little to the south of west. There was no perceptible wind at the moment. The witness's attention was attracted first to the freak the wind
had in the grass, and the next moment to a hissing noise caused by the cinders passing through the air. The larger ones were considerably imbedded in the earth, so much as only to show a small part of it, while the smaller ones were about one-half buried. The witness noticed at the time that the ground (where he afterwards picked up the cinders) showed signs of warmth, as there was quite a steam or fog at that particular point. He thought it singular, as the ground had been very cold previously. According to the article the "scoria" was in rounded inflated pieces, like what have been called volcanic bombs, the exterior being glassy, and the interior very cellular. They are little over an inch in the longest diameter.

This description of the 1857 fall event has some resemblance with the 2003 Elma event.

However there are events without any rainy weather, and moreover sometimes even without reports of "black clouds, etc." which hints that there was nothing special in the weather from the eyewitness viewpoint (so it was not even mentioned). Here are several examples below.

- The Igast event took place in Livonia (nowadays Estonia) in the place Igast ( Ihaste ) which is the suburban neighbourhood the city of Tartu nowadays. Here is the description of this remarkable event adapted from (Wichmann, 1913).

In the afternoon of the 17th of May 1855 at about 6 o'clock a lady, Miss Beckmann, standing on the stairs outside a country-seat 2 miles N.E. from Walk in Livonia, suddenly saw between the lime-trees on that spot a dazzling luminous phenomenon, at about 6 or 9 feet above the ground, whilst she heard at the same time a tremendous clap. At about the same time the proprietor, Mr. Fr. Schultz, found himself in a rather high field situated at a distance of 1 km from the county-seat. He, likewise, heard in an entirely cloudless sky a violent detonation, so strong that his saddle-horse and another horse, drawing a barrow, were frightened and threatened to bolt.

Assuming that a falling meteorite had exploded, the chemist L. Bornwasser immediately made an investigation on the ground in the neighbourhood of the above-mentioned lime-trees, and collected two handfuls of peculiar mineral's fragments which he supposed to proceed from that meteorite.

Some additional info is in (O'Keefe, 1960): "The fall was unusually well observed, there being two witnesses; one faced the point of fall and was only 50 feet away at the instant of fall. The fall itself was unmistakable, with a flash and a detonation. The pieces were found loose on the grass under some linden trees that had been cut by the detonation".

Since that time numerous investigations of the fragments were conducted. Already results of the first ones pointed that the object was not a meteorite (some researchers pointed that it resembles a slag), i.e. placed it into pseudo-meteorite category. The last research (known to the author) was published in 1966 (Lowman and O'Keefe, 1966) and confirmed the terrestrial origin of the Igast object. The Igast object is marked as a pseudo meteorite in the Meteoritical Bulletin Database which has been constructed and is maintained by the Nomenclature Committee of the Meteoritical Society ( https://www.lpi.usra.edu/meteor/about.php ).

- A fall of another piece of a remarkable stone in Europe was described in (Baker, 1959). Here is its a little bit adapted description. The event took place at Halle-Heide at 2.30 p.m. on 14th August, 1883. Some people heard a humming noise in the branches of a nearby tree, and noticed an accompanying appearance of light. In the steaming soil of a nearby spot, they reputedly found a slag-like, black stone that had loosened the soil to a depth of 10 cms. It was still too hot to touch and was lifted from the hole with two sticks. This glass was subsequently shown to be basic glass similar to the artificial glass spheres from Kuttenberg in Czechoslovakia. In thin section, it is light yellowish-green, and contains sharply defined crystals of leucite, pyroxene, plagioclase, apatite, olivine and melilite.

The fallen object is absent in the Meteoritical Bulletin Database which has been constructed and is maintained by the Nomenclature Committee of the Meteoritical Society ( https://www.lpi.usra.edu/meteor/about.php ).

- The 1872 fireballs fall in a lightship event (Allen, 1873). A letter, addressed

by the Secretary of the Corporation of the Trinity House to the President of the Royal Society, states that at 2 a.m. Nov.13, 1872 a meteor burst over the "Seven Stones" light-vessel, moored about 9.5 miles E. by N. of the Scilly Islands. It has been reported that the watch were struck senseless for a short period, seeing nothing before the shock, but on recovery they observed:

*"...balls of fire like large stars were falling in the water like splendid fireworks, and that the decks were covered with cinders, which crushed under the sailors' feet as they walked. The superintendent reports that the men say there was a very decided smell of brimstone, but add that they did not mention that until he asked them. There is reason to fear that the cinders were all washed off the decks by the rain and sea before daylight; and it happened also unfortunately that the men did not think to observe the compasses."*

Words "washed off the decks by the rain" hint that it rained at the time of the event or some time later. The latter seems to be more probable, as otherwise the rain soon would be mentioned in the description. Anyway it looks like there was nothing peculiar in the weather as it was not mentioned in the report. Remarkably that the "cinders" were transported into the sea for at least 15 km.

- The 1925 pseudo-meteorite fall in Sweden. Unfortunately, publications on studies of pseudo-meteorites are rarely found in the scientific literature. One of the few exceptions is article (Cross, 1947). It describes several inexplicable cases of the fall of strange "meteorites." One of them occurred on April 11, 1925 in Sweden (Cross, 1947).

A witness ("a respected and reliable man") informed researchers that he had seen the falling "white ball" in the sky, which then broke against the ground. Further investigation showed that the fragments composed of almost pure calcium carbonate. Also small fragments of calcareous shell were scattered through it. Some of the fragments had a peculiar gloss, which researchers had never seen "in natural limestones or in burnt or otherwise prepared stones".

The fallen object is named as Bleckenstad and marked as a pseudo meteorite in the Meteoritical Bulletin Database which has been constructed and is maintained by the Nomenclature Committee of the Meteoritical Society
( https://www.lpi.usra.edu/meteor/about.php ).

In (Cross, 1947), several other discoveries of limestone's pieces with a melted surface are also given.

- The May 10, 1931 "copper" fall in Colorado, USA. The following description of the event is based on (Nininger, 1943, 1972). It took place on the sunny afternoon. An eyewitness ( Mr. W. H. Foster ) was hoeing in his garden at Eaton, Colorado, when he was attracted by a humming sound, not unlike that of a stray bullet, which seemed to come from the northern sky. The sound seemed louder, as if approaching him. Foster leaned on his hoe handle and listened; this seemed a long time for a bullet to whine.  He scanned the sky but saw nothing. Soon he felt the air blast in his face as an object whizzed past and struck with a thud about 2 meters south and a little west of where he stood.  Looking down he saw where the sun-baked crust of the soil had been broken up, and projecting from the moist dirt thus exposed was a small, bright bit of metal. Foster pondered for a moment, then stooped to pick up the object, "burning" his fingers, he said later, as he did so. His neighbor superintendent John C. Casey testified that the Foster's finger showed a fresh burn half way between its tip and the first joint (Nininger, 1943). There were no airplanes overhead on that morning (Nininger, 1943). Remarkably that "By careful questioning and by re-enactment of the scene, it was evident that Mr. Foster had heard the whining noise of the "bullet" for a minute or two minutes before its landing" (Nininger, 1972).

According to (Buseck et al., 1973) the Eaton object (weight about 30 g.) contains roughly 66 percent Cu, 33 percent Zn and <0.1 percent Ni. The major phases and inclusions of Eaton closely resemble those in commercial yellow brass.

The fallen object is named as Eaton `and` marked as a pseudo meteorite in the Meteoritical Bulletin Database which has been constructed and is maintained by the Nomenclature Committee of the Meteoritical Society ( https://www.lpi.usra.edu/meteor/about.php ). Remarkably that the noise associated with the event was heard for a minute or two minutes before the object's landing (Nininger, 1972).

Next events were reported in 1930s in relation with so called "tektites problem". The author of the reports considered these falls as tektites falls (australites) in modern times in Australia.

- The alleged australites falls in 1930s. Description of the first reported event was published in (Simpson, 1935). An eyewitness with several other men were working in a field in the summer 1932-1933, when something flew past him with a hissing noise "like a piece of shrapnel", and struck the ploughed earth with an audible thud only a few feet away. He noticed a small hole in the ground, and underneath it at a depth of about 30 cm he found the "australite". Although it was obtained within about three minutes of its fall, it did not appear to the eyewitness to possess a temperature differing from that of the soil in which it was embedded. According to

Simpson E.S. the stone looks like "typical ellipsoidal australite, weighing 31 grammes".

Description of the second "australite" event is taken from (Simpson, 1939). Early in 1935 (the exact date is forgotten) Mr. F. Hanson was working on a gravel tennis court in Kathleen Street, North Cottesloe (near the city of Perth), when he heard, at about 10 or 11 a.m., a thud on the surface of the court. He observed a thin cloud of dust rising about 9 meters away, and found evidence of some object having penetrated the surface there. Digging on the spot immediately, he found beneath 7.5 cm of laterite gravel (used to surface the court) and a further 30 cm of loose sandy soil, a rounded stone weighing about 140 g, which was still too hot to hold in the hand. In (Bowley, 1945) a couple of interesting details were added: there was a flash of light and a smell.

The data on the following research of the fallen objects is a bit controversial (Baker, 1959).
Anyway both objects are absent in the Meteoritical Bulletin Database which has been constructed and is maintained by the Nomenclature Committee of the Meteoritical Society ( https://www.lpi.usra.edu/meteor/about.php ).

- The 1977 Iowa, USA "molten iron" event. This event occurred at Council Bluffs, Iowa, on December 17, 1977 and is described in (Vallee, 1998; Wilson, 2001), and is briefly presented below with little adaptation.

Several residents of the town observed a bright flash at 7.45 p.m. The flash was followed by flames 2.5 - 3 m high. When the witnesses reached the scene of the event, they found a large area on the northern city limits, covered with a mass of molten metal that glowed red-orange and had ignited the grass. Its dimensions were about 1.2 m by 2.7 m. There was a secondary burn area ~8 m away, measuring about 0.6 by 1.2 m. Also many metal spherules were found scattered about the area.

There was almost a dozen witnesses of the event in all. Two residents of the town Council Bluffs saw an object that crashed to the ground on the northern city limits. The initial witnesses were Kenny Drake and his wife Carol, and Kenny's 12-year old nephew Randy James. Two other witnesses, Mike Moore and his wife Criss, reported seeing a hovering red object with lights as they crossed 16th street on their way downtown along Broadway avenue. Criss reported "a big round thing hovering in the sky below the tree tops. It was hovering. It wasn't moving." She added that she saw red lights around the perimeter of the object, blinking in sequence. A middle-aged couple who saw the event spoke to the investigators by telephone, stating that they had seen a bright red object rocket to the ground near Big Lake but refused to be identified.

Secondary witnesses who observed the metal were Jack E. Moore, assistant fire chief (who took the 911 call from Kenny Drake), police officer Dennis Murphy and Robert E. Allen, who had served in the Air Force and wrote a weekly astronomy column for a local newspaper. Mr. Moore stated that the center of the metal mass was too hot to touch when he arrived on the scene about 8 p.m., only 15 minutes after the initial incident, and that it remained so for about an hour.

Investigation proceeded as follows: Measurements taken at the impact point by Robert Allen indicate the object was traveling from the Southwest to the Northeast. Samples of the object were sent to the Ames Laboratory at the Iowa State University, and others were taken to the Griffin Pipe Products Company. The material was determined to be carbon steel, probably man-made, of a type common in

manufacturing.  The following four hypotheses were examined and then rejected by the investigators:

a)  An unknown person poured molten metal on the ground as a hoax;

b)  An unknown  person created  molten metal as a hoax by using thermite and ordinary metal;

c)  The material came from an aircraft;

d) The event was due to a meteoritic impact, or space debris.

So the origin of the event remained unidentified.

- The 1983 fall in then-USSR at the town of Türkmenabat nowadays (Turkmenistan). Description of the event is taken from (Florensky et al., 2016).  The event took place on August 11-12, 1983 at about 21.30 hours, along the route near the railroad leading to the automobile bridge over the Amu Darya river.

An eyewitness on August 11-12, 1983 at about 21.30 hours saw a hissing object similar to a red flare. The object flew from the southwest to the northeast and fell on a cotton field. In the morning the eyewitness found a crater in the middle of a cotton field - a depression similar to the nest of a large bird, about 20 cm in depth and a diameter of about 60 cm. Around the crater, cotton bushes were felled to the outside. There was in the crater a glass stone similar to obsidian with a melted surface and traces of "blowing off", weighing 340 g, size about 100 × 70 mm, density about 2.81 g/cm$^3$. There were two more fragments 7 cm long and a little thinner than 10 mm thick. The crater was covered with black dust, like very small balls.

The event was registered in the Committee on Meteorites of the Soviet Academy of Sciences, but the fallen object is absent in the Meteoritical Bulletin Database which has been constructed and is maintained by the Nomenclature Committee of the Meteoritical Society ( https://www.lpi.usra.edu/meteor/about.php ).

- The Jan. 29, 1986 event near the town of Dalnegorsk, USSR. Soon after the event various interpretations were proposed - from meteorite fall, and spy-balloon self-destruction, and to "aliens". Some "aliens"-gravitated info can be read here: https://en.wikipedia.org/wiki/Height_611_UFO_incident

In scientific circles the most popular is idea of a kind of a ball-lightning. Fortunately in 1988 the place was researched by a group of scientists under the leadership of experienced geologist Vladimir Salnikov. Several years later some results of the research were presented (Salnikov, 1990). Here is just a couple of points from (Salnikov, 1990):

a) Unaffected soil samples have silica content within 90-93%, and in the impact place 84-89%.

b) Results of paleomagnetic analysis of siliceous rocks from the impact area show that the effect exerted by the impact led: 1) to a local increase of magnetic susceptibility, which can occur by inserting of magnetic material or in case of a change in the iron oxide pigment structures; 2) to increase the residual magnetization and change in the polar plane of magnetization, which is probably caused by the action of the magnetizing field; 3) to changing the angle of the radial projection of the vector up to 90°. If the interpretation model is correct, then the minimum field strength is $10^3$ - $10^4$ A/m, temperature in the center is not less than 400-500° C, and the diameter of the object is not more than 1.0-1.5 m.

- The March 25, 1993 tektite-obsidian-like glass fall in Japan (Shima et al., 1994). Despite that not many details of its fall are known, the author (i.e. A. Ol'khovatov) decided to include this event, as there are some resemblance with the above-mentioned "australites" fall events in 1930s.

Here is from (Shima et al., 1994). On March 25, 1993, around 2pm, a black glassy stone, 122.2 g, fell on the balcony of the office at 7th floor of the tall building in the center of Kyoto City. The office is only the tall building, so it cannot be the results of somebody's joke. The external appearance is a duplicate of the "Tektite". Determination of cosmogenic radioactive nuclides, petrographical characteristics, major and trace elements and other research was performed. As the results, it is concluded that the glass is neither like as tektite nor similar to terrestrial natural glasses such as obsidian. It could be originated in the moon. It was temporaly named as Kyotite. That was what the stone researchers thought.

Remarkably, that a spacebody infall delivering at least 122 g on the ground must be associated with a rather bright bolide. But reports about such bolide are absent.

The fallen object is absent in the Meteoritical Bulletin Database which has been constructed and is maintained by the Nomenclature Committee of the Meteoritical Society ( https://www.lpi.usra.edu/meteor/about.php ).

- The 1994 pseudo-meteorite fall in Spain (Martínez-Frías et al., 1999). In 1994 a moving car and its driver, on a highway in southern Madrid (Getafe), were struck by a falling rock. Eighty-one additional fragments (total weight 55.926 kg) were later recovered, which all pointed towards a meteorite fall. A study of the composition and isotopic data of this object revealed an ultrarefractory material displaying an unusual chemical make-up which differs from any known meteorites, and hints at its terrestrial origin. According to (Martínez-Frías et al., 1999) the only artificial material which presents some compositional similarities to the Getafe rock is a specific type of primary steelmaking slag: namely Electric Arc Furnace (EAF) slags. These EAF slags are crystalline solids, with the textural and chemical appearance of igneous rock, which have a high density ( 2.4 g/cm$^3$ — approximately half that of the Getafe rock ) and a compositional variability depending on the proportion in which components are artificially mixed. The Getafe rock is officially catalogued as a pseudo-meteorite in the meteorite collection of the National Museum of Natural Science, Madrid (Spain) (Muñoz-Espadas et al., 2002). The fallen object is marked as a pseudo-meteorite in the Meteoritical Bulletin Database which has been constructed and is maintained by the Nomenclature Committee of the Meteoritical Society
( https://www.lpi.usra.edu/meteor/about.php ).

So there are a lot falls resembling the 2003 Elma event even without a thunderstorm. The latter fact makes search for possible physical mechanism even more complicated. However some hints exist which could help associate these events with some other natural phenomena. A ball-lightning was mentioned in this article early in association with a thunderstorm. But a ball-lightning can exist without a thunderstorm too (Turner, 2002):

*"Ball lightning is almost invariably observed during thundery weather, though not necessarily during a storm. Similar, though possibly not identical, phenomena have been repeatedly observed in specific geographical areas and in association with earthquakes, also in some of the accounts of*
*unidentified flying objects (UFOs) given by aircraft pilots. In these cases, however, stormy conditions are not necessary. It may be significant that observations made in fine weather often refer to a somewhat larger and much longer-lived phenomenon than is typical for ball lightning. It could be that the two kinds of ball differ in their detailed electrochemistry, the prolonged existence of the fair weather form resulting in part from a markedly different, fair weather, space charge."*

The author of this article prefers to call high-speed ball-lightnings (so often resembling meteors and bolides of extraterrestrial origin) as "geophysical meteors" (Ol'khovatov, 2000). As it is known that a ball-lightning can carry some substance (see early in this article), then the 2003 Elma event can be called as an example of a geophysical meteors.

Some other events are known which could carry substance in the above-given examples. William Corliss in a chapter of his book (Corliss, 1977) "Whirlwinds and dust devils" wrote that "Whirlwinds and dust devils, however, may be violent on occasion, sometimes appearing very suddenly with a peculiar detonation or rumbling sound", and devoted a section called "Explosive onset of whirlwinds" to several such events. Here is probably one of the most remarkable (Corliss picked up the text from "London Times", July 5. 1842).

*"Wednesday forenoon [June 29] a phenomenon of most rare and extraordinary character was observed in the immediate neighborhood of Cupar. About half past 12 o'clock, whilst the sky was clear, and the air, as it had been throughout the morning, perfectly calm, a girl employed in tramping clothes in a tub in the piece of ground above the town called the common, heard a loud and sharp report overhead, succeeded by a gust of wind of most extraordinary vehemence, and only of a few moments duration. On looking round, she observed the whole of the clothes, sheets, &c. , lying within a line of certain breadth, stretching across the green, several hundred yards distant; another portion of the articles, however, consisting of a quantity of curtains, and a number of smaller articles, were carried upwards to an immense height, so as to be almost lost to the eye, and gradually disappeared*

*altogether from sight in a south-eastern direction and have not yet been heard of. At the moment of the report which preceeded the wind, the cattle in the neighboring meadow were observed running about in an affrighted state, and for some time afterwards they continued cowering together in evident terror. The violence of the wind was such that a woman, who at the time was holding a blanket, found herself unable to retain it in fear of being carried along with it! It is remarkable that, while even the heaviest articles were being stripped off a belt, as it were, running across the green, and while the loops of several sheets which were pinned down*
*and snapped, light articles lying loose on both sides of the holt were never moved from their position."*

In the Corliss's book there is another event, which probably also could better understand the effect of "transportation". Here it is ( original text was published in "Symons's Meteorological Magazine") from (Corliss, 1977):

"A NARROW SQUALL
*Clark, J. Edmund; Symons's Meteorological Magazine, 49:52, 1914.*

*On 19th July, 1913, a narrow strip of wind $1^1/_2$ ft. wide, drove through a hedge lifting newly cut hay 60 yards high, proceeded through a tree with great violence, blowing off the leaves of a hedge and tree, and upset a hay-cock in an adjoining field. It lasted some 10 minutes and was accompanied by great noise. I was informed by an old inhabitant such occurrences were frequent at Coquet Head in the Cheviots. I doubt they are all as narrow as this was. (Symons's Monthly Meteorological Magazine, 49:52, 1914)"*

Remarkably that an experienced "taiga traveler" and researcher of the 1908 Tunguska event - Konstantin Kokhanov ( who searched for the alleged Tunguska spacebody remnants for several decades in the regions surrounding the Tunguska event area) wrote in his blog (http://parfirich.kohanov.com/blog/?p=114, translated from Russian by A. Ol'khovatov):

*"The author of the article did not have to observe such tornadoes in Siberia, with catastrophic consequences, but he observed three times the formation and action of small tornado-like vortices.*

*The first time he encountered this phenomenon was in 1973 on the Bolshaya Erema river, as soon as he got into a boat and made several strokes with an oar near the river bank downstream the Bur rapids. When the boat was half a meter from the left bank, something hissed from its side and fell into the water, right in front of its bow. Almost instantly, a foam dome about 1.5 meters high and about 2.5 meters in diameter formed. Konstantin Kokhanov flinched in surprise, thinking that a deafening explosion would follow, but a small swirling column of water only formed above the foam dome. At first, he emitted a slight whistle, which then with increasing volume, together with the foam dome, rushed towards the large channel of the Bur rapids. <...>*

*The second time he met with a similar "tornado" or natural phenomenon in 1979 on the Altyb River, ten kilometers from its mouth, from the left bank of the Bolshaya Erema river. That whirlwind was smaller, but more noisy, as if chomping at unsuccessful successive starts of the outboard motor.*

*"Tornado" unexpectedly formed somewhere on the side of the opposite left bank of the Altyb, twirled in the bank's bushes, then lifting, as if sucking water into itself to a height of a little more than a meter, rushed across the river, and passed from him about ten meters downstream and got lost in the valley of two low and gentle hills.*

*For the third time, Konstantin Kokhanov saw a similar "tornado", being on the bank of the Nizhnyaya Tunguska river, 2 kilometers upstream the village of Erema, in the company of fifteen people from the village who invited him there to celebrate some local holiday. He did not even see it at the beginning, but heard a cannon shot and, turning towards the river, he saw in the middle of the riverbed a water dome formed there, at least ten meters in diameter and at least three meters high, which quickly fell off and did not make another sound. Having asked one of the local residents, who also turned towards the river, - what it was - Kokhanov heard in response that this phenomenon is nothing particularly rare here - just a "whirlwind"."*

In his another blog post (http://parfirich.kohanov.com/blog/?p=3950) K. Kokhanov added some interesting aspect regarding the third event (translated from Russian by A. Ol'khovatov):

*"At the same time, no sleeve or funnel of the tornado that had lowered onto the river was not visible. An interesting reaction to this phenomenon was a local resident who stood with him. When Konstantin Kokhanov asked what it was, he replied - nothing special, just a "whirlwind". The rest of the fishermen sitting on the bank did not even turn around to see what it rattled on the river."*

Regarding Kokhanov's information, it is interesting to mention another natural phenomenon that was observed on the Russian Far North and received the name Tonge (Alekseeva, 1985). The author of the book (Alekseeva, 1985) - a geophysicist wrote about popular hero of the stories of the inhabitants living in the area of Antipayuta (a settlement in the Taz Estuary region).

Around Antipayuta, brigades of Nenets reindeer herders roam the tundra with their herds. It was these people that Tonge began to worry about. Tonge appeared with the first twilight after the polar day, when the ground had just begun to be covered with snow. To some witnesses Tonge appeared in the form of a thin white pillar going up into the air. Another saw a similar pillar while on the move, when it suddenly appeared between two sledges. To the third it appeared to be a monstrously "tall white man" with luminous eyes. Someone noticed a chain of his footprints with several "fingers", always stretching somewhere into the distance.

Tonge appeared, apparently at night. It began to "play" with a tent (chum): It climbed on its wall, slipped sticks under it, threw inside the tent through the smoke hole various small objects that it picked up in the tundra - usually stones and sticks, but one day an old kerosene lamp fell from above, and another time a bottle of ink for a fountain pen. Sometimes Tonge threw away the deer skin, which covers the entrance to the chum, and then one could notice his "eyes" - white, green or red. The man who looked out of the chum did not see anyone, but one day someone leaned out and received a blow on the back with a stick. Occasionally there were sounds like the neigh of a horse. In the morning, on the wall of the chum covered with snow, the "letters" of Tonge were visible.

Panic broke out among the elderly and children. The reindeer herders' foreman told how he and a group of other men jumped out of the chum during the Tonge raid. Standing around the chum and holding hands, they began to converge, wanting to grab Tonge. At the same time, they did not see or hear anything special, but all the time it seemed to them that they were being shot at by bullets. Catch, of course, failed. Even a commission left the neighboring settlement to investigate the reasons for the anxiety and unnecessary migrations, as well as, if possible, calm the population down.

Alekseeva incline to think that Tonge appears during the years of high solar activity.

But unfortunately due to limited data it is hard to say surely. It would be interesting to understand whether there is some relation between Tonge and another rare phenomenon - so called "Very Low Auroras" - there is a section on them in (Corliss, 1977). Unfortunately the existing data is too sparsely for any reliable conclusion.

The feeling "being shot at by bullets" in the above-given case hints that possibly Tonge was associated with high electric field in the atmosphere. Here is an example how this field can manifest, taken from a geologist's story (Popov, 2010, translated from Russian by A. Ol'khovatov):

*"I'll tell you about one interesting atmospheric phenomenon with which I had to face in the Suntar-Khayata mountains.*

*From afar came the weak thunder rolls. The light rain, which had been drizzling since early morning, stopped. Examining rock outcrops, making the necessary notes. Suddenly I pay attention to the annoying sound above my head, remind the buzzing of a large fly. Mechanically I wave my hand away, I slap on my cap - the sound does not disappear. Unexpectedly, the mustache moved, and immediately a sharp and strong whistle of an invisible "arrow" was heard about flying over the ground about ten meters from me, in the direction across my route. I froze in one place, and the invisible "arrows" the flight of which could only be monitored by sound, have already flown around me on three sides. "The air is electrified" - came the guess, and I looked apprehensively at my iron hammer.*

*Having chosen the moment of calm, I carefully lower the hammer to the ground, I move away from him and hide myself behind a small stone. "Arrows" first*

*in a flock, then one by one, they again began to frolic in the air in search of a target. After 5-10 minutes a gust of wind flew in, fell from the sky snow pellets, and electrical discharges stopped."*

But probably the most intriguing phenomenon occurred in a factory (compilation of the info is collected by William J. Beaty on his web-site http://amasci.com/weird/unusual/e-wall.html ).
The factory's workers encountered a strange "invisible wall" in the area under a fast-moving sheet of electrically charged polypropelene film. This "invisible wall" was strong enough to prevent humans from passing through. A person near this "wall" was unable to turn, and so had to walk backwards to retreat from it. Whatever its physical mechanism, the phenomenon hints that an object sometimes can be "trapped" and probably transported in such conditions.

So there are some hints that the phenomena discussed can be related with peculiar manifestations of electricity in the atmosphere. But hints are not proofs, so let's continue with phenomenological approach.

Now let's consider whether a kind of a whirlwind could heat a transported object. The answer (based on observations) is - "at least some of them can". Indeed William Corliss in his book (Corliss, 2001) has a section "Whirlwinds of Fire and Smoke". Such whirlwinds seems to be rather rare, anyway here is one example taken from (Anonymous, 1869), and which was reproduced by W. Corliss in 2001 (Corliss, 2001):

*"Out in Cheatham county about noon on Wednesday — a remarkably hot day — on the farm of Ed. Sharp, five miles from Ashland, a sort of whirlwind came along over the neighbouring woods, taking up small branches and leaves of trees and burning*

*them in a sort of flaming cylinder that travelled at the rate of about five miles an hour, developing size as it travelled. It passed directly over the spot where a team of horses were feeding and singed their manes and tails up to the roots; it then swept towards the house, taking a stack of hay in its course. It seemed to increase in heat as it went, and by the time it reached the house it immediately fired the shingles from end to end of the building, so that in ten minutes the whole dwelling was wrapped in flames. The tall column of travelling caloric then continued its course over a wheat field that had been recently cradled, setting fire to all the stacks that happened to be in its course. Passing from the field, its path lay over a stretch of woods which reached the river. The green leaves on the trees were crisped to a cinder for a breadth of 20 yards, in a straight line to the Cumberland. When the "pillar of fire" reached the water, it suddenly changed its route down the river, raising a column of steam which went up to the clouds for about half-a-mile, when it finally died out. Not less than 200 people witnessed this strangest of strange phenomena, and all of them tell substantially the same story about it. The farmer, Sharp, was left houseless by the devouring element, and his two horses were so affected that no good is expected to be got out of them in future. Several withered trees in the woods through which it passed were set on fire, and continue burning still . -- Nashville (Tennessee) Press.*

To close discussion of whirlwinds, let us consider one event, where presence of whirlwind is not evident, but anyway the event was associated with stones throwing up in the air. This tragic event occurred in Italy in 1761 is presented below from ( Anonymous, 1762 ). The text is a little bit adapted for modern spelling, and also description of wounds is mostly omitted due to ethical reasons. Under the name Morand (mentioned in there) probably was an outstanding French surgeon Sauveur François Morand.

*"A woman of the village of Bonne - Vallie, near Ventimillia, aged about 37 years, was*

*returning with four of her companions, from the forest of Montenere, each being loaded with a bundle of small sticks and leaves, which they had been gathering. As soon as they arrived at a place called Gargan, this woman, two of her companions being before, and two behind her, suddenly cried out with great vehemence, and immediately fell down with her face towards the ground. The person that was nearest to her observed nothing more than usual, except a little dust that rose round her, and a flight motion in some little stones that lay upon the spot ; they all ran immediately to her assistance, but they found her quite dead ; her cloaths, and even her shoes were cut, or rather torn into slips, and scattered at the distance of five or six feet round the body, so that they were obliged to wrap her up in a cloth, in order to carry her to the village.*

*Upon inspedling the body, the eyes appeared fixed and livid; there was a wound on the left side of the os frontis, which left the pericranium bare, and there were also many superficial scratches upon the face in strait lines. <...> ...the muscles of the buttock and thigh were almost carried away, and what is yet more astonishing, notwithstanding this loss of flesh, which could not be less than six pounds, there was not the least drop of blood to be seen upon the spot where the accident happened, nor the least fragment of the flesh that had been torn away.*

*It was supposed that this poor woman was killed by the eruption of a subterraneous vapour, which issued from the ground directly under her; a conjecture which seems the more probable, as, in the summit of the mountain Montenere, there are two chinks, from which smoke frequently issues, and at the foot of the mountain there is a sulphureous spring.*
*<...>*
*This most extraordinary relation was communicated by M. Morand, to the royal academy of sciences in Paris, by whom it has been made public."*

Absence of the flesh and the blood mentioned above hints that it was carried away by the "burst". The proposed explanation that this poor woman was killed by the eruption of a subterraneous vapour looks unlikely, as no fissure, etc. was reported. Also if even such event happened, then (from description of the wounds) it follows that it would be rather high-speed outburst accompanied with powerful sounds, and other effects, also other persons would be affected by it too.

Remarkably there was in some way similar event occurred in Hungary in 1989. The description of the event is taken from (Meaden, 1990). Original investigation was made by Dr George Egely of the Central Institute of Physics, Budapest. The event occurred on 25 May 1989, near near a village 109 kilometers from Budapest. The victim was a 27-year old engineer. The man had stopped his car and walked to the edge of a field about ten meters distant to urinate. Suddenly his wife who had remained behind in the car saw that he was surrounded by a blue light. He opened his arms wide and fell to the ground. His wife ran to him, noticing that one of his tennis shoes had been torn off. Although it looked hopeless she tried to help him but soon after she was able to stop a passing bus. Amazingly, the bus was filled with medical doctors returning from a meeting; they immediately pronounced that the man was dead. At the autopsy a hole was found in the man's heel where the shoe had been. The lungs were torn and damaged, and the stomach and belly were carbonized. At the time of the event the sky was overcast. The couple had earlier passed through a thunderstorm when 50 kilometres away, but at the site of the tragedy it was dry and rainless.

History preserved for us another remarkable event, which is in some way similar to the above-mentioned (the main difference is a sound of an explosion). Its description is taken from (Brydone, 1787). The event took place near Coldstream

(England). There was several lightnings with thunders on July 19, 1785 between 12 am and 1 pm at a considerable distance from the place. Soon after a powerful report was heard. The following investigations "on hot traces" by Mr. Brydone discovered the following.

Two carts carrying coal were about to enter the top of a small hill when the sound of a powerful explosion occurred. A driver of the carriage moving behind the leading one saw as the driver in the leading cart fell to the ground, and with them both of their horses and the leading cart fell too. All they were dead. At the time of the explosion the second driver saw no fire, felt nothing unusual. The dead carter's body was burned in various places, clothes were torn. The hair was much singed over the greatest parts of the horse's bodies, and it was the most perceptible on the belly and legs.

The left shaft of the leading cart was broken; and splinters had been thrown off in many places, particularly where the timber of the cart was connected by nails, or cramps of iron. Many pieces of the coal were thrown out to a considerable distance, all round the cart; and some of them have the appearance of coal which had lain some time on a fire.

About 4.5 feet behind each wheel of the cart, an odd appearance in the ground was observed - a circular hole of about twenty inches in diameter, the center of which was exactly in the track of each wheel. The earth was torn up, as if by violent blows of a pick-axe, and the small stones and dust were scattered on each side of the road. The tracks of the wheels were strongly marked in the dust, both behind and before these holes, but were completely obliterated for upwards of a 1.5 foot on these spots.

The surface of the iron on the cart's wheels, to the length of about three inches, and the whole breadth of the wheel, had become of a bluish colour, had entirely lost its polish and smoothness, and had the appearance of drops incompletely formed on its surface. To ascertain whether these marks were occasioned by the explosion which had torn up the ground, the cart was pushed back on the same tracks which it had produced on the road; and it was found, that the marks of fusion are exactly at the center of each of the holes; and that, at the instant of the explosion, the iron of the wheels had been sunk deep in the dust. The broken earth still emitted a smell something like that of ether. The ground was remarkably dry, and of a gravelly soil. But the wheels and the horse's legs and bellies must had been apparently wet due to river crossing just before climbing the small hill. Mr. Brydone also examined all horse's shoes, but could not perceive the least mark on any of them. A shepherd (being at some distance from the carts) was looking at the two carts, but he saw no lightning, no fire, but observed the dust to rise at the place.

Mr. Brydone presents several other phenomena happened on that day. The shepherd belonging to the farm of Lennel-hill was in a neighbouring field, tending his flock, when he observed a lamb drop down; and said, he felt at the same time as if fire had passed over his face though the lightning and claps of thunder were then at a great distance from him. He ran up immediately, but found the lamb quite dead; though the moment before it appeared to be in perfect health. This happened about a quarter of an hour before the explosion which killed the carter; and it was not above 300 yards distant from the spot. He was only a few yards from the lamb when it fell down. The earth was not torn up, nor did he observe any dust rise.

Two fishers were standing in the middle of the Tweed river, fishing for salmon with the rod, when they suddenly heard a loud noise; and, turning round to see from whence it came, they found themselves caught in a violent whirlwind, which felt sultry and hot, and almost prevented them from breathing. It was not without much difficulty they could reach the bank. However it lasted but a very short time, and was succeeded by a perfect calm. This happened about an hour before the explosion.

A woman, making hay near the banks of the river, fell suddenly to the ground, and called out to her companions, that she had received a violent blow on the foot, and could not imagine from whence it came.

A respectable witness said, that, while walking in his garden, a little before the explosion, he several times felt a sensible tremor in the ground.

Remarkably that after the explosion the clouds began to dissipate and there were no more either thunder or a lightning.

Mr. Brydone added some other data to his report "as it may possibly have had some influence on it." The drought was very great till the 22d of July, when it rained a little; and this was repeated, though in small quantities, and generally accompanied by high winds, till Thursday the 27th, when it blew the most violent tempest he ever remembered in this country. On Aug.11 there was an earthquake.

In the author's opinion the original report by Patrick Brydone, Esq. F.R.S. is worth to read. The author expresses deep respect and admiration for the investigation and to its creator.

Let us demonstrate how knowledge of the above-given events probably could help to understand (on the phenomenological level) some peculiarities of the 1908 Tunguska event. For example, why there is a boulder near the Tunguska epicenter, which according to some research fell on the ground, but detailed investigation of its geochemical data points that it is a terrestrial rock (Haack et al., 2016).

Interestingly that there is at least one specific report about a fall of a large stone during Tunguska. Evenk woman Nastya Dzhenkul ( the widow of Ivan Maksimovich Dzhenkul ) informed in November 1959 (Vasil'ev et al., 1981) (translated from Russian by A. Ol'khovatov) that:

*"Her father and grandfather lived at that time (in 1908) on the Khushma River. The weather was good, suddenly it began to rain, a strong wind rose, and dragged away the birch bark tent. A large stone fell, as big as a tent, jumped two or three times, and then drowned in a swamp. The stone was shiny, black, fell with a terrible sound - u - u -u -u."*

Remarkably that Innokentii Suslov in a letter to a journalist wrote in 1961 about his participation in the Evenk's congress in 1926
( http://tunguska.tsc.ru/ru/science/mat/oche/31-60/d-041/ , translated from Russian by A. Ol'khovatov):

*"The article "Agdy" contains detailed stories of the Tungus (Evenks) who flew above the forest, saw a burning forest, smoke under them, survived the horror of the roar, blows and were on the verge of losing their minds. They spoke in detail about the details of the disaster.*

*All this was written in the well-established Tungus-Russian Angara dialect, which contributes to a brighter view of the reader about the disaster. But the most interesting thing there, in my opinion, is a detailed description of Munzhen (general meeting of voters on the modern constitution), a historical meeting of Evenks on the*

*split of the Chunya River, where at the beginning of June 26, the first lower bodies of Soviet power were elected - tribal councils (two ) and with them tribal courts. At the end of this meeting, as an experiment of anti-religious propaganda, I was able to raise the issue of the place where the meteorite fell.*

*The experience was a success. He challenged the Evenks to frankness at such an authoritative meeting. As a result, all the stories of individual victims (who flew) were confirmed here... "*

Unfortunately to the author's knowledge, the article "Agdy" was never published, and the "flights" are absent in (known to the author) Tunguska accounts presented by Suslov. The reason for this is unclear, but maybe poor health of I. Suslov's wife (who typed his notes) was involved. Also possibly it was because to some reviewers these Evenks accounts seemed "too incredible", and Suslov (having problems with health) later decided not to insist and to exclude them. Such reaction happens regarding some accounts of Tunguska - see, for example (Ol'khovatov, 2020), despite that Evenks are natural hunters and pathfinders (their life depends on this). Moreover Suslov considered the accounts to be very reliable also because they were collected during Munzhen, when telling a lie was considered to be a serious misconduct.

Please pay attention that the reported Evenk's flying seems to occur some time after starting fires, as the forestfires seem to be rather developed at these moments. It is in agreement with Evenk's "on the ground" accounts some of which point that a windstorm continued for a relatively long time (at least comparing with an air-shock from an alleged spaceimpact).

Regarding the reported Evenk's flying it is reasonable to consider a possibility of tornadic like wind in the area. In 1928 a prominent Soviet meteorologist Boris Multanovskii said about the possibility of cyclonic phenomena in the Tunguska epicenter which could cause the devastation discovered by L. Kulik. However, Multanovskii pointed out, due to the absence of meteorological stations in 1908 in the area, there is no weather data at the disposal of the Weather Bureau.

It is possible to add that meteorological logs from small weather stations (some of them were closest to the Tunguska epicenter) were hard to obtain. So for many years most of the weather analysis was done basing on the remote weather stations and eyewitnesses who were far from the Tunguska epicenter. Such analysis hints to more-or-less good weather near the Tunguska epicenter near the time of the explosion. However since late 1950s conducted large-scale eyewitness survey revealed that some of the Evenks (which lived not far from the Tunguska epicenter ) said about rain, etc. (see, for example, Nastya Dzhenkul - above). Moreover some details of the forestfire's developments hint that the forest was wet, or became wet after the starting fire. In addition in 2008 a new work appeared (Gorbatenko, 2008), which says (among other things) that the meteorological situation was rather complicated (TKT is the alleged Tunguska space body, translated from Russian by A. Ol'khovatov):

"*Judging by the changes in air temperature and pressure at stations located around
the place where the TKT fell and during the fall, it can be assumed that over the territory of the south-west of the Eastern (and partly in the southeast of Western) Siberia there was an extensive cyclone. Its warm front follows passed through the settlements: Kansk, Yeniseisk, Kezhma. The center of the cyclone was located east of Krasnoyarsk, in the Kansk region, with a pressure in the center of about 735 mm Hg (980 mb).*

*In addition, analyzing the air temperature values (Fig. 2) at stations close to the area of interest to us, it is safe to state that in the period from 06/30/1908 to 07/01/1908 above*
*the territory of interest to us passed a warm front and for 0.5-1 days the territory was in a warm*
*sector of the cyclone, and therefore at this time there could be stratus clouds and a general deterioration visibility. A little later it was replaced by a cold front with cumulus clouds and thunderstorms."*

The latter work takes into account data from the small stations, which increases reliability of results. It would be nice to use computer simulation of the meteorological situation basing on all data avaiable for more accurate analysis. Remarkably, for example, a weather log of the Kirensk meteostation (situated about 491 km to the SE from the Tunguska epicenter ) marked in the afternoon of June 30 a strong wind. For comparison, the previously strong wind was noted by this meteorological station in 1908 only on March 31, and next time - December 9th. So June 30, 1908 was a peculiar day in the region indeed.

Regarding the above-mentioned aspects of Tunguska, probably it would by reasonable to add that some interesting formations were discovered several dozens km from the Tunguska epicenter. The first one is unusual explosive structure discovered by a geologist in the late 1970s (Dorofeev, 2008). Here is from (Dorofeev, 2008):

*"One of my reconnaissance routes passed along a left side of a brook valley with latitudinal stretch, a left confluent of the Yuzhnaya Chunya river (several tens of kilometers to the northeast form the epicenter of the Tunguska explosion). The*

*route run through the field of development of grey-colorful loosely consolidated silt-psammitic tuffites of early Triassic age, very typical for the upper part of a stratigraphic section of the Korvunchana suite. An isoline of relief on the other side of the valley made a distinctive bend at one of the sections of my route. Such situation in locality must be connected with a terrace. But river terraces for upper reaches of small streams are extremely atypical. One could most likely expect to meet there an outcrop of a bedded body of magmatic rocks – dolerite or metasomatic rocks with quartz-carbonate composition. However I found a turfy terrace made from a breccia – a result of a powerful explosion. The breccia consisted of debris of basaltic lava from 1 to 10–15 cm in size. I was astonished by a variety of colors from grey-green to violet-brown, with predominance of various hues of red. The debris were consolidated by a chemical cement – zeolite (heulandite), and this was also unusual (because analcime is typical for cements of tuffs at that location). The breccia represented a dense and strong rock, which, judging by the terrace, lied subhorizontally. Father, the route run in the field of development of these breccias consolidating a gently sloping watershed. At a distance of 2–3 km I found a hole similar to a crater which had one wall destroyed by erosion, so that in view from above the structure had a shape of a sector covering 300–330 degrees. Its upper diameter was 100–150 m, the diameter at the bottom was about 50 m, and the depth was about some tens of meters. I saw the same breccia at the slopes and bottom of the crater but the debris reached 1–1.5 m in size there. A spring welled out at the bottom, and a streamlet flowed out through a narrow mouth of the destroyed wall and fell into the main valley. Along the periphery of the structure, I found several outcrops of quartz-carbonate metasomatites, which were bow-shaped in view from above, with a curvature center coinciding with the bottom of the hole. Later, during office studies, analyzing geophysical data, I found that the field of the breccia spreading is associated with a magnetic anomaly of high intensity. <...>*
*(1) I could not determine, even roughly, the area of breccia spreading,*

*however, I think that it is longer than 3 km in linear dimension. A lower boundary is early Triassic, but the upper one is not determined because there are no dated sediments which would intersect the breccia, and the thickness of modern (Holocene) sediments is minor. (2) The explosion, to all appearances, was of near-surface character, because rocks of pre-volcanic substrate are absent in breccias structure, as well as tuffites embracing them. (3) A clear negative shape of relief was evident. There is a rule in geomorphology: if a structure is older, it is worse expressed in relief, and vise versa. Therefore, the suggestion about a relatively young age of the structure does not contradict the actual situation. (4) Connection between the Tunguska "meteorite" and the explosive structure is unclear, however, it seems quite possible because the finding is extremely uncommon."*

Interesting iron formations were discovered under the surface lay of soil in 2013 at the distance 40 km east (and slightly north) from the Tunguska epicenter (Dmitriev, 2016). 12 samples of the iron were collected, total weight 96 kilograms. One sample weighing 3.5 kg was different from the rest, an alloy of iron and dark green glass. Some samples are a calcium aluminosilicate. The dimensions of the investigated area: 600 m x 2200 m, The samples were at the depth 15 - 45 cm.

The weight of the discovered samples is from 1 kg to 42 kg. Samples have splinter-torn structure. The shape is convex like a broken lens from glasses. Sample thickness is from 5 cm to 18 cm. All samples have melting crust. In places of splitting the melting crust is much thinner. Most of the samples consist of native iron with the inclusion various minerals. The iron content is 40-60%. Completely amorphous small glasses up to 1.2 cm in various shapes, uneven stripes and droplets, also fragments of basalt structure and fine pumice were discovered on the surface of some samples. The samples are not melted in the central internal part. Their mineral composition: metallic native iron, magnetite, quartz, orthoclase, albite, wollastonite. Unfortunately these formatioms are poorly researched. All hopes are for future expeditions.

It looks like something really intriguing happened in Tunguska, but detailed discussion of the 1908 Tunguska event is beyond the scope of this article.

## 4. Conclusion

Having deal with poorly understood and poorly investigated phenomena, the author of this article decided to use phenomenological approach and concentrate on facts. The facts demonstrate that events similar to the 2003 Elma event are reported from time to time. There are hints of their possible connections with a ball-lightning and some type of a whirlwind - so in general with some processes in the atmosphere which are not completely understood. Anyway knowledge of these phenomena allows to look from the different angle at some unusual aspects of the 1908 Tunguska event. And these peculiar events mean that our planet is full of miracles still.

**ACKNOWLEDGEMENTS**

The author wants to thank the many people who helped him to work on this article, and special gratitude to his mother - Ol'khovatova Olga Leonidovna (unfortunately she didn't live long enough to see this article published...), without her moral and other diverse support this article would hardly have been written.